\documentclass[prb, twocolumn,showpacs,superscriptaddress, preprintnumbers,amsmath,amssymb]{revtex4}

\usepackage{graphicx}
\usepackage{dcolumn}
\usepackage{bm}
\usepackage[dvipsnames]{color}

\begin{document}


\title{Photoemission evidence for a Mott-Hubbard metal-insulator transition in VO$_2$}

\author{R.~Eguchi}
\email[Electronic mail: ]{ritsuko@spring8.or.jp}
\affiliation{Soft X-ray Spectroscopy Laboratory, RIKEN SPring-8 Center, Sayo-cho, Sayo-gun, Hyogo 679-5148, Japan}

\author{M.~Taguchi}
\author{M.~Matsunami}
\author{K.~Horiba}
\author{K.~Yamamoto}
\author{Y.~Ishida}
\author{A.~Chainani}
\author{Y.~Takata}
\affiliation{Soft X-ray Spectroscopy Laboratory, RIKEN SPring-8 Center, Sayo-cho, Sayo-gun, Hyogo 679-5148, Japan}

\author{M.~Yabashi}
\affiliation{Coherent X-ray Optics Laboratory, RIKEN SPring-8 Center, Sayo-cho, Sayo-gun, Hyogo 679-5148, Japan}
\affiliation{JASRI/SPring-8, Sayo-cho, Sayo-gun, Hyogo 679-5198, Japan}

\author{D.~Miwa}
 \altaffiliation{Deceased.}
\author{Y.~Nishino}
\author{K.~Tamasaku}
\affiliation{Coherent X-ray Optics Laboratory, RIKEN SPring-8 Center, Sayo-cho, Sayo-gun, Hyogo 679-5148, Japan}

\author{T.~Ishikawa}
\affiliation{Coherent X-ray Optics Laboratory, RIKEN SPring-8 Center, Sayo-cho, Sayo-gun, Hyogo 679-5148, Japan}
\affiliation{JASRI/SPring-8, Sayo-cho, Sayo-gun, Hyogo 679-5198, Japan}

\author{Y. Senba}
\author{H. Ohashi}
\affiliation{JASRI/SPring-8, Sayo-cho, Sayo-gun, Hyogo 679-5198, Japan}

\author{Y.~Muraoka}
\author{Z.~Hiroi}
\affiliation{Institute for Solid State Physics, University of Tokyo, Kashiwanoha, Kashiwa, Chiba 277-8581, Japan}

\author{S.~Shin}
\affiliation{Soft X-ray Spectroscopy Laboratory, RIKEN SPring-8 Center, Sayo-cho, Sayo-gun, Hyogo 679-5148, Japan}
\affiliation{Institute for Solid State Physics, University of Tokyo, Kashiwanoha, Kashiwa, Chiba 277-8581, Japan}

\date{\today}

\begin{abstract}

The temperature ($T$) dependent metal-insulator transition (MIT) in VO$_2$ is investigated using bulk sensitive hard x-ray ($\sim$ 8 keV) valence band, core level, and V 2$p$-3$d$ resonant photoemission spectroscopy (PES). The valence band and core level spectra are compared with full-multiplet cluster model calculations including a coherent screening channel. Across the MIT, V 3$d$ spectral weight transfer from the coherent ($d^1\underbar{\it {C}}$ final) states at Fermi level to the incoherent ($d^{0}$+$d^1\underbar{\it {L}}$ final) states, corresponding to the lower Hubbard band, lead to gap-formation. The spectral shape changes in V 1$s$ and V 2$p$ core levels as well as the valence band are nicely reproduced from a cluster model calculations, providing electronic structure parameters. Resonant-PES finds that the $d^1\underbar{\it{L}}$ states resonate across the V 2$p$-3$d$ threshold in addition to the $d^{0}$ and $d^1\underbar{\it {C}}$ states. The results support a Mott-Hubbard transition picture for the first order MIT in VO$_2$. 
\end{abstract}


\maketitle

VO$_2$, a $d^1$ electron system, exhibits a sharp first-order metal-insulator transition (MIT) as a function of temperature ($T$), at $T_{MI}$ = 340 K. \cite{morin}  The high-$T$ metal phase has a rutile ($R$) structure, while the low-$T$ insulating phase has a monoclinic ($M_1$) structure with zig-zag type pairing of V atoms along the c-axis. \cite{pouget} Magnetically, the metallic $R$ phase shows enhanced susceptibility ($\chi$) with an effective mass $m^{*}/m$ $\sim$ 6, while the insulating $M_1$ phase is non-magnetic. Consequently, VO$_2$ has attracted enormous attention in terms of a Mott-Hubbard (MH) correlation-induced versus a structural Peierls-type MIT.

Goodenough proposed that the lowest energy $t_{2g}$ states split into the $d_{\parallel}$ band and the $\pi^{*}$ bands. In the $R$ phase, the $d_{\parallel}$ band overlaps the $\pi^{*}$ band. \cite{good} In the $M_1$ phase, the zig-zag type pairing consists of (i) a tilting of the V-V pairs which lifts the $\pi^{*}$ band above the Fermi level ($E\rm{_F}$), and (ii) 
the $d_{\parallel}$ band opens up a  gap because of the splitting into a filled bonding and empty antibonding band caused by V-V pairing. Zylberstejn and Mott attributed the MIT to an on-site Hubbard type Coulomb energy $U$ in the $d_{\parallel}$ band. \cite{zyl} Pouget {\it et al\/}. showed that pure VO$_2$ under  uniaxial stress, and Cr-doped VO$_2$,  exhibit an insulating ($M_2$) phase \cite{pouget, rice} in which, half of the V atoms form pairs, and the other half form zig-zag chains that behave as spin-1/2 Heisenberg chains in NMR and EPR experiments. \cite{pouget2} Another intermediate phase [triclinic ($T_r$)] connecting $M_2$ $\rightarrow$ $M_1$, showed a bonding or dimerization transition of the Heisenberg chain V atoms, with a concomitant decrease of $\chi$. Thus, it was concluded that the $M_1$ insulating phase of VO$_2$, as well as the $M_2$ and $T_r$ phases, are MH insulators. 
In an alternative picture, VO$_2$ was explained in terms of a Peierls insulating phase due to the structural phase transition. The electron-phonon interaction was considered to be important from Raman scattering \cite{sriva} and x-ray diffraction studies. \cite{mcwhan} An ab-initio molecular dynamics study in the local-density approximation (LDA) described VO$_2$ as a charge-ordered band (Peierls) insulator, \cite{wentz} but a genuine gap was not obtained. Also, the linear resistivity of metallic VO$_2$, analysed in terms of an electron-phonon effect in a Fermi liquid, gives an invalid mean free path. \cite{wentz} The LDA approach for VO$_2$ was discussed in comparison with NbO$_2$ and MoO$_2$, but does not provide a gap in the density of states (DOS). \cite{eyert} Combining O 1$s$ x-ray absorption spectroscopy (XAS) and ultraviolet(UV)-photoemission spectroscopy (PES), the value of the $d_{\parallel}$ band splitting was estimated to be 2.6 eV, whereas the LDA calculates it to be 2.0 eV. The authors concluded that the structural distortion is dominantly responsible for the $d_{\parallel}$ band splitting in the insulating phase. \cite{abbate}

The gap in the DOS, and suppressed moment, are correctly obtained in the cluster dynamical mean field theory (c-DMFT) calculations of Biermann {\it et al\/}., showing the crucial role of strong Coulomb interactions and structural distortions in VO$_2$. \cite{bier} 
The c-DMFT gives an isotropic metal, with the $d^1$ electron occupancy: 0.36 in the $d_{\parallel}$ ($\equiv$ $a_{1g}$) band and 0.32 each in the $\pi^{*}$ ($\equiv$ $e_{g}$ doublet) bands. In the insulating phase, occupancies change to 0.8 in the $d_{\parallel}$ ($\equiv$ $a_{1g}$) band and $\sim$ 0.1 in each of the $\pi^{*}$ ($\equiv$ $e_{g}$ doublet) bands. Biermann {\it et al\/}. conclude that VO$_2$ exhibits {\it a correlation assisted Peierls transition}. Haverkort {\it et al\/}., using polarization dependent XAS, have shown the importance of orbital occupancy switching in the V 3$d$ states across the MIT, \cite{Haverkort} in very good agreement with the occupancies calculated in a cluster model, as well as c-DMFT results. Haverkort {\it et al\/}. conclude an {\it orbital assisted collaborative Mott-Peierls transition} for VO$_{2}$.

VO$_2$ has provided other important results: an electric-field-induced MIT \cite{Kim} and a photo-induced MIT. \cite{Cava1} The electric-field-induced MIT, in the absence of a structural transition, favors a MH transition. From the ultrafast ($<$ psec.) time scale of the photo-induced MIT in VO$_2$, \cite{Cava1} the study concluded that the accompanying structural transition may not be thermally initiated. 
In VO$_2$/TiO$_2$:Nb thin films, the photo-induced electron-hole pairs result in a reduced resistivity by 10$^{-3}$ times in the insulating phase and a surface photovoltage effect. \cite{mura} We recently reported the photo-induced electronic structure changes using soft x-ray(SX)-PES, showing the combined effects of rigid band behavior and correlation effects. \cite{egu} Several PES studies \cite{Sawatzky, shin, goe, oka, egu, Koethe} have addressed the electronic structure changes across the $T$-dependent MIT in VO$_2$. Early work discussed the role of strong correlations versus structural effects in VO$_{2}$. \cite{Sawatzky, shin, goe}  A $T$-dependent UV-PES  study with an analysis of surface and bulk components, suggested the importance of both electron-electron and electron-phonon interactions for the electronic structure of VO$_2$. \cite{oka} 
More recently, infrared spectroscopy and nano-scale microscopy was used to reveal a divergent quasi-particle mass of the metallic carriers an approaching the insulating phase. \cite{Qazilbash} The insulating phase of VO$_2$ was classified as a Mott insulator with charge ordering. Thus, we felt it important to study the MIT in VO$_2$ using bulk sensitive hard x-ray(HX)-PES, complemented by full-multiplet cluster calculations and V 2$p$-3$d$ resonant-PES. Recent studies using HX-PES have shown new features in core level spectra which could be consistently explained in terms of well-screened peaks. \cite{taguchi, Kamakura, horiba2} A study on valence band and core levels of V$_2$O$_3$ using HX-PES has shown a clear correlation between the well-screened peak and the coherent peak at $E\rm{_F}$. \cite{Panaccione} Theoretical studies have also shown the importance of coherent screening \cite{Cornaglia} as well as non-local screening \cite{Veenendaal} in core level spectra.

In this work, we study VO$_2$ thin films using HX-PES ($h\nu$ = 7937 eV).  
Since the mean free path (MFP) of electrons using HX-PES is significantly longer (6 - 10 nm) than that using UV or SX-PES ($<$ 2 nm), \cite{seah} it provides the bulk electronic structure.  However, the photoionization cross sections (PICS) become very small ($\sim$ 10$^{-5}$ times) compared to UV or SX photons and severely limit valence band studies. In order to enhance signal intensity, and since HX attenuation lengths are much longer than relevant MFPs, we adopted a grazing incidence geometry. \cite{takata} This necessitated the use of VO$_2$ thin films grown on a single crystal surface of TiO$_2$:Nb, allowing us an incidence angle of  about 1$^{\circ}$.
We measure the valence band, the V 1$s$, V 2$p$ and O 1$s$ core level spectra as a function of $T$. Resonant-PES across the V 2$p$-3$d$ threshold, and a cluster model calculation of the valence band and the core level V 1$s$, V 2$p$ spectra provide a consistent picture of electronic parameters for VO$_2$.

A 10-nm-thick VO$_2$ film was epitaxially grown on the (001) surface of 0.05 wt\% Nb-doped TiO$_2$ single crystal substrates using the pulsed laser deposition technique, as described elsewhere. \cite{mura2} The tensile strain results in a first order hysteretic transition just like bulk VO$_2$, but at a slightly lower temperature of $T_{MI}$ $\sim$ 292 K.
HX-PES and SX-PES experiments were carried out at undulator beam line BL29XUL, SPring-8 (Ref.\ \onlinecite{ishikawa}) using a Gammadata-Scienta R4000-10 KV spectrometer \cite{takata} and BL17SU, SPring-8 using a Gammadata-Scienta SES2002 spectrometer. \cite{HoribaOhashi} The energy resolution was set to 200 $\sim$ 250 meV. The measurements were carried out in a vacuum of 5 $\times$ 10$^{-8}$ Pa, at 270 and 320 K. 
The $E\rm{_F}$ of VO$_2$ was referenced to that of a gold film evaporated onto the sample holder in all measurements. 

\begin{figure}
\begin{center}
\includegraphics[scale=0.67]{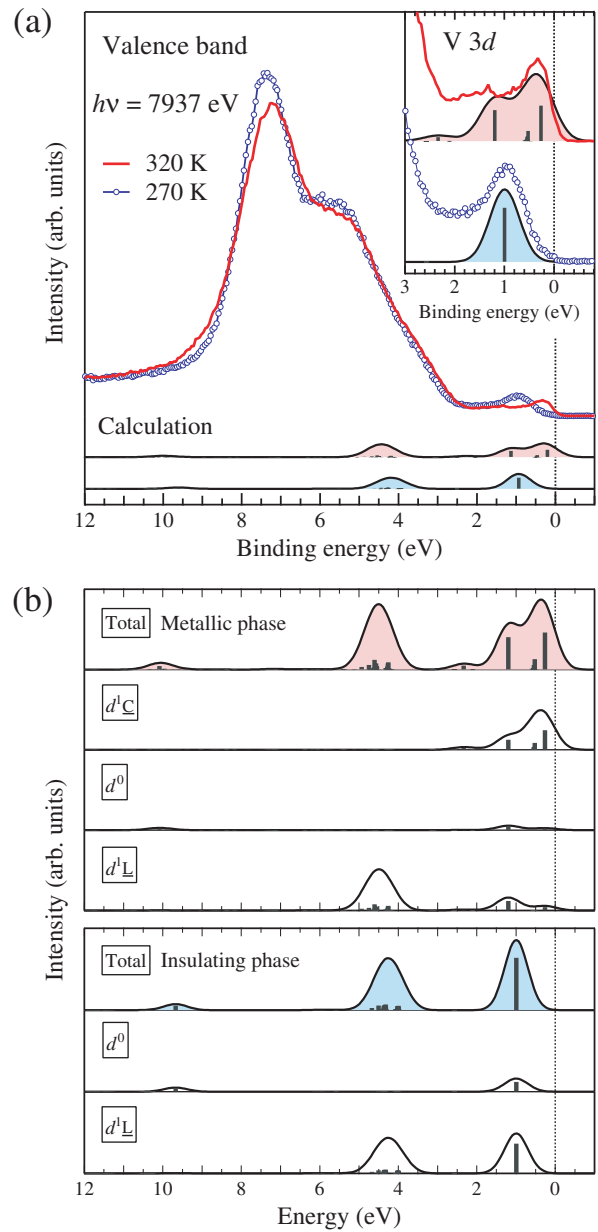}
\end{center}
\caption{(Color online) (a) Valence-band spectra measured across the MIT using  hard x-ray ($h\nu$ = 7937 eV), compared with the cluster model calculation. Inset shows an expanded view (0-3 eV) of the V 3$d$ states near $E\rm{_F}$. (b) The main final state configuration spectra in the metallic phase and the insulating phase.}
\end{figure}

\begin{figure*}
\begin{center}
\includegraphics[scale=0.68]{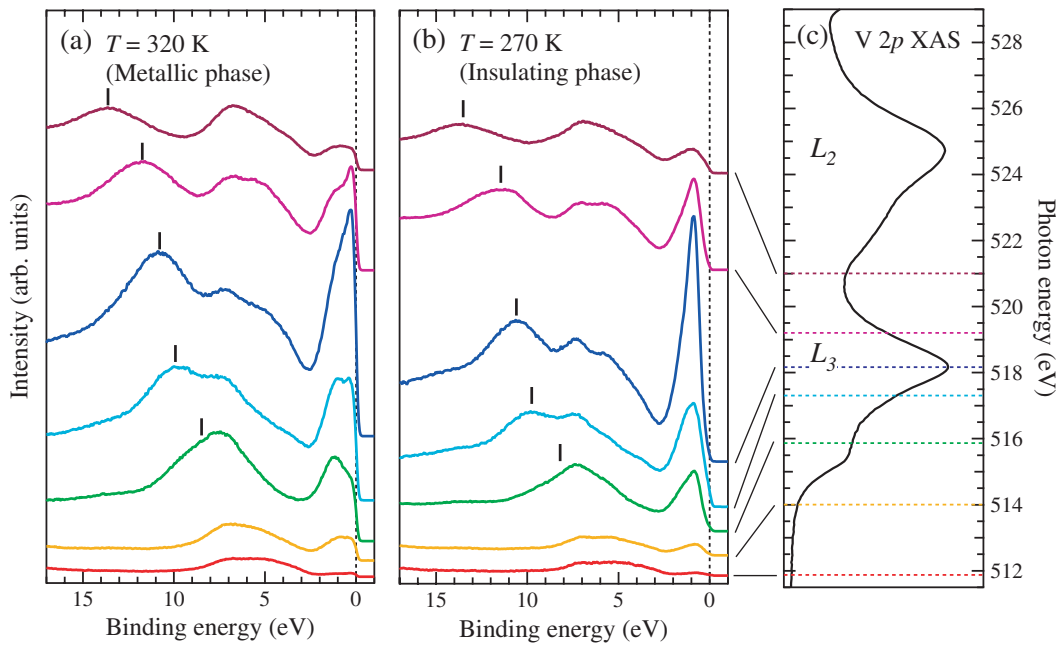}
\end{center}
\caption{(Color online) Resonant-PES across the V 2$p$-3$d$ threshold in (a) the metallic phase and (b) the insulating phase. (c) V 2$p$ XAS spectra.}
\end{figure*}

Figure 1(a) shows the raw data of the entire valence-band of VO$_2$ measured using HX-PES at 320 and 270 K, normalized for area under the curve. The valence band structure consists of the dominantly V 3$d$ band near $E\rm{_F}$ [0-2.5 eV binding energy] and a broad band usually considered as the dominantly O 2$p$ band between 3 to 10 eV in previous results. \cite{shin, oka} Based on PICS and a comparison with SX-PES results, \cite{Koethe} the spectral shape of the 7.5 eV feature suggest that the V 4$s$ contribution overlaps the O 2$p$ states.  
The energy position and width of the so called O 2$p$ band change slightly across the MIT, consistent with SX-PES results. \cite{Koethe} 
Figure 1(a) inset shows an expanded view (0-3 eV) of the V 3$d$ states near $E\rm{_F}$, which provides a direct picture of the gap formation in VO$_2$. The 320-K spectrum shows a peak centered at 0.3 eV and a clear Fermi edge, indicating the metallic state, and another broad weak feature centered at about 1.5 eV. These features correspond to the coherent band at $E\rm{_F}$ and the incoherent (lower Hubbard) band, respectively. The larger intensity in the near $E\rm{_F}$ DOS using HX-PES is attributed to the bulk sensitive V 3$d$ states, since the PICS of V 3$d$ is higher than O 2$p$ at high-photon (-kinetic) energy. The 270-K spectrum near $E\rm{_F}$ shows large changes compared to the 320-K spectrum, with a peak centered at 1 eV and negligible intensity at $E\rm{_F}$,  indicative of an energy gap of about 0.2 eV.

Moreover, in order to investigate the origin of these structures in detail, we carried out calculations for a VO$_6$ cluster model in $O_h$ symmetry, as described in earlier work. \cite{taguchi, Kruger} For simplicity, we assume a octahedral symmetry throughout. Since the crystal field effect on peak intensities and positions in photoemission is small even for the (generally dominant) cubic contribution, it is clear that the neglected lower symmetry terms would not change the spectral shapes in any significant way as far as photoemission is concerned. A study of lower symmetry is beyond the scope of this work. In addition to the usual charge transfer energy, $\Delta$ from the O 2$p$ ligand band to the upper Hubbard band, the charge transfer energy from the coherent band to the upper Hubbard band is defined as $\Delta{^*}$. The ground state is described by a linear combination of following configurations: 3$d^1$, 3$d^2\underbar{\it {L}}$, 3$d^3\underbar{\it {L}}^2$, 3$d^2\underbar{\it {C}}$, 3$d^3\underbar{\it {LC}}$, 3$d^3\underbar{\it {C}}^2$. The parameters were set as follows: $\Delta$ = 4.0 eV, $\Delta{^*}$ = 0.2 eV, $U_{dd}$ = 4.5 eV, 10$Dq$ = 1.2 eV, $V$ = 2.4 eV, $R_v$ = 0.8 eV, in addition, $U_{dc}$(1$s$) = 8.0 eV, $U_{dc}$(2$p$) = 6.5 eV, $R_c$(1$s$) = 0.9 eV, and $R_c$(2$p$) = 0.8 eV for the core level calculations [Fig.\ 3(b) and (c)]. Theoretical calculations for transition metal impurities in a solid give a $U_{dd}$ $\sim$ 7.5 eV for V, which gets reduced by typically $\sim$ 60 \% in an oxide. \cite{HameraChen} The calculations reproduce the experimental spectra in the metallic and the insulating phases. All the parameters for the calculated spectra for the metallic and insulating phases are equal, except for the hybridization between the central V atom 3$d$ orbitals and the coherent band, $V^*$. $V^*$ = 0.48 eV for the metallic phase, and it was set to $V^*$ = 0 eV for the insulating phase. This confirms the importance of the coherent screening channel in reproducing the metallic phase of VO$_2$.  The discrete levels obtained from the calculations were broadened by a Gaussian function of full width at half maximum $\Delta E$ = 0.35 eV.

\begin{figure}
\begin{center}
\includegraphics[scale=0.75]{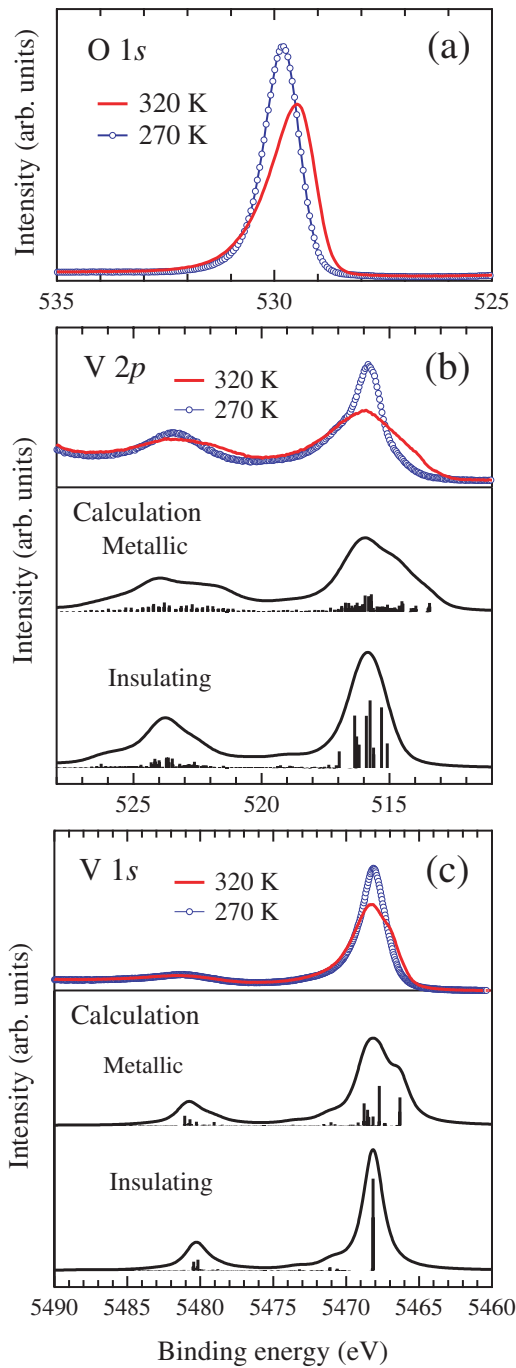}
\end{center}
\caption{(Color online) (a) O 1$s$, (b) V 2$p$, and (c) V 1$s$ core level spectra measured across the MIT using  hard x-ray ($h\nu$ = 7937 eV), compared with the cluster calculations for the metallic phase and the insulating phase.}
\end{figure}

Figure 1(b) shows the main final state configuration (3$d^1\underbar{\it {C}}$, 3$d^0$, and 3$d^1\underbar{\it {L}}$) spectra in the metallic and the insulating phase. The total spectra are also shown in Fig.\ 1(a). 
In the metallic phase, the coherent screening state 3$d^1\underbar{\it{C}}$ is dominant near Fermi level. On the other hand, the locally screening 3$d^ 0$ state has a large intensity at a binding energy of $\sim$ 1 eV with weak contribution from $d^1\underbar{\it {L}}$ states. In insulating phase, the coherent screening states vanish and the locally screened 3$d^ 0$ with $d^1\underbar{\it {L}}$ admixture states remain around 1 eV. This feature corresponds to the effective lower Hubbard band, nicely matching the incoherent band in PES spectrum. The main contribution of $d^1\underbar{\it {L}}$ state appears around 4 eV in the metallic and insulating phases. We have checked that the ground state configuration is a mixed character [MH+ charge transfer (CT)] state. The difference compared to an earlier study \cite{Mossanek} stems from a dominantly MH description in the present study compared to dominantly CT description in Ref.\ \onlinecite{Mossanek}. Specifically, Ref.\ \onlinecite{Mossanek} uses $\Delta$ = 2.0 eV and $U$ = 4.5 eV, while in present case, we have used $\Delta$ = 4.0 eV and $U$ = 4.5 eV.

The single-site DMFT (Ref.\ \onlinecite{lie}) and the c-DMFT (Ref.\ \onlinecite{bier}) calculations, both indicate large changes in charge redistribution across the MIT. The c-DMFT calculation predicts a weak intensity lower Hubbard band in the insulating phase at $\sim$ 1.8 eV, with the quasi-particle origin feature shifted to 0.8 eV, for a $U$ = 4 eV and $J$ = 0.68 eV. However, the HX-PES shows no feature at $\sim$ 1.8 eV, but a single peak at 1.0 eV 
that is attributed to 3$d^ 0$ state (the lower Hubbard band) in the cluster model calculation result. 
The c-DMFT study points out that a lower $U$ (e.g. 2 eV) also stabilizes a gap, provided $J$ is also small. However, it probably implies that the quasi-particle feature will move to even smaller energies than 0.8 eV. Instead, a c-DMFT with a larger $U$ and $J$ is probably more appropriate for VO$_2$. As has been discussed, while single-site DMFT results can also give a gap with a large $U$ ($\sim$ 5 eV), it also gives local moment. While additional calculations can provide quantitatively accurate comparisons, both, the single site and c-DMFT would require a $U$ $>$ 4 eV.  
Thus we have set $U_{dd}$ = 4.5 eV as a reasonable value in present calculations.

Using XAS and resonant-PES across the V 2$p$-3$d$ threshold, we show the enhanced V 3$d$ electronic structure. It resonates due to interference between the direct channel ($p^6d^{n} + h\nu \rightarrow p^6d^{n-1} + e$) and the photoabsorption channel ($p^6d^n + h\nu \rightarrow p^5d^{n+1} \rightarrow p^6d^{n-1} + e$) excitations. Resonant-PES is well-known for probing this behavior in correlated oxides like NiO, CuO, etc. and $f$-electron systems. \cite{oh,allen} Using photon energies marked in the XAS profile of Fig.\ 2(c), we measured valence band PES in the metallic [Fig.\ 2(a)] and the insulating phase [Fig.\ 2(b)]. 
These spectra are normalized to the scan time and the incident photon flux.
 The off-resonance spectrum is very similar to the SX-PES spectrum. \cite{Sawatzky,egu, Koethe} The spectra show a typical resonance enhancement of the 0-2 eV feature as a function of  $h\nu$, with a maximum at $h\nu$ = 518.2 eV ($L_3$ peak in XAS). This confirms 3$d$ electron character of the 0-2 eV feature. Moreover, the higher-binding feature of O 2$p$ band gets enhanced at a lower energy ($\sim$ 514 eV), and is further enhanced at 516 eV. At higher $h\nu$, a strong intensity Auger feature shows up at higher binding energy and tracks the increase in $h\nu$ [marked in Fig.\ 2(a) and 2(b)].  
The valence band calculations suggest the existence of the main $d^1\underbar{\it {L}}$ state overlapping the O 2$p$ band. It resonates due to interference between the direct channel ($d^2\underbar{\it {L}} + h\nu \rightarrow d^1\underbar{\it {L}}+ e$) and the photoabsorption channel ($p^6d^2\underbar{\it {L}} + h\nu \rightarrow p^5d^3\underbar{\it {L}} \rightarrow p^6d^1\underbar{\it {L}} + e$). Note that the $d^1\underbar{\it {L}}$ final state position is not clearly fixed in the present calculation, because  the present cluster calculations give a discrete level for the $d^1\underbar{\it {L}}$, unlike actual $d^1\underbar{\it {L}}$ states having a band like energy distribution. The Auger features are hardly observed in the 0-2 eV resonant feature, very similar to the case of Ti$_2$O$_3$ resonant-PES, another 3$d^1$ system. \cite{ParkTezuka}

Figure 3(a) shows the O 1$s$ core level HX-PES spectra at 320 and 270 K. The O 1$s$ core level shows clear spectral shape change; a symmetric peak in the insulating phase transforms to an asymmetric Doniach-{\v{S}}unji{\'{c}} line shape in the metal phase. No contamination feature at higher binding energy to the main peak is observed.  Figure 3(b) shows the V 2$p$ core level spectra at 320 and 270 K. The V 2$p_{3/2}$ and V 2$p_{1/2}$ main peaks are positioned at $\sim$ 516 eV and $\sim$ 524 eV. The V 2$p$ spectra show remarkable changes across the MIT: the peak shape is sharp in the insulating phase, but shows a shoulder structure at low binding energy in the metal phase.  In addition, the V 1$s$ core level spectra at 320 and 270 K show very similar changes like the V 2$p$ spectra, as shown in Fig.\ 3(c). The V 1$s$ main peak is positioned at 5468 eV, with a weak satellite at 5481 eV in the insulating phase. The main peak and the satellite feature get broadened in the metallic phase, with clear additional structure at lower binding energies.

We also carried out cluster calculations using the same set of parameters as for the valence band calculations. The calculations [Fig.\ 3(b) and 3(c)] nicely reproduce the V 2$p$ and V 1$s$ core level experimental spectra. The shoulder structure appears in the metallic spectrum, whereas the insulating phase spectrum has no shoulder structure in the V 1$s$ as well as V 2$p_{1/2}$ and V 2$p_{3/2}$ core levels.  The results confirm that the screening channel from the coherent band at $E\rm{_F}$ is responsible for the shoulder structure. This result is slightly different from HX-PES results reported in V$_2$O$_3$ (Refs.\ \onlinecite{taguchi} and \onlinecite{Kamakura}) and La$_{1-x}$Sr$_x$MnO$_3$, \cite{horiba2} which show a clear additional peak derived from the coherent screening in both HX-PES core level spectra and the cluster model calculations. Since the additional peak position is affected by the magnitude of the parameters, $\Delta$,  $\Delta{^*}$, and $U_{dc}$ in the calculation, \cite{taguchi} the additional peak appear well-separated from the main peak and the ligand screening (2$p^5$3$d^{n}\underline{L}$) peak in V$_2$O$_3$ and La$_{1-x}$Sr$_x$MnO$_3$, resulting in a distinguishable peak structure. In contrast, since the $\Delta$ is relatively small compared to $U_{dc}$ for VO$_2$, and unlike V$_2$O$_3$ and La$_{1-x}$Sr$_x$MnO$_3$, the coherent screening state is positioned near the ligand screening state, resulting in a broad shoulder structure. The overall calculation results thus reproduce the core level and valence band PES spectra and provide consistent electronic structure parameters with a $U$ $\gtrsim$ $\Delta$.

In conclusion, we have performed $T$-dependent valence band and core level HX-PES of VO$_2$. The valence band spectra show gap formation and weight transfer from the coherent state (the quasi-particle band, $d^{2}\underbar{\it {C}} \rightarrow d^{1}\underbar{\it {C}}$ character) to a mixed incoherent state [the lower Hubbard band, ($d^{1} \rightarrow d^{0}$) + ($d^{2}\underbar{\it {L}} \rightarrow d^{1}\underbar{\it {L}}$) character], supporting a MH transition picture. Resonant-PES across the V 2$p$-3$d$ threshold finds that the $d^1\underbar{\it {L}}$ state, which overlaps the O 2$p$ band,  resonates as well as $d^{0}$ and $d^{1}\underbar{\it {C}}$. The spectral shape changes in V 1$s$ and V 2$p$ core levels as well as the valence band are nicely reproduced from a cluster model, providing a consistent set of electronic structure parameters.

The HX-PES experiments reported here have benefited tremendously from the efforts of Dr.\ D. Miwa of the coherent x-ray optics laboratory RIKEN/SPring-8, Japan and we dedicate this work to him.





\end{document}